\documentclass[english,aps,prm,superscriptaddress,reprint,longbibliograhy]{revtex4-2}

\usepackage[T1]{fontenc}
\usepackage[latin9]{inputenc}
\usepackage{amsmath}
\usepackage{amssymb}
\usepackage{amsfonts}
\usepackage{bbold}
\usepackage{tikz}
\usepackage{xfp}

\usepackage{color}

\usepackage{graphicx}
\usepackage{dcolumn}
\usepackage{bm}
\usepackage{commath}
\usepackage{upgreek}
\usepackage{siunitx}
\usepackage{physics}
\usepackage[mathlines]{lineno}
\usepackage[english]{babel}

\usepackage[bottom=2cm,top=2cm, left=1.5cm, right=1.5cm]{geometry}
\usepackage{booktabs}
\colorlet{linkequation}{blue}
\usepackage[colorlinks=true,linkcolor=cyan,citecolor=red,urlcolor=cyan]{hyperref}
\usepackage[normalem]{ulem}
\hypersetup{pdftitle={Role of sublattice volume ratio and Gilbert damping in THz field-derivative-torque-induced nonlinear magnetization dynamics}, pdfauthor={Arpita Dutta, Pratyay Mukherjee, Swosti P. Sarangi, Somasree Bhattacharjee, Christian Tzschaschel, Shovon Pal and Ritwik Mondal}}

\begin{document}

\title{Role of material-dependent properties in THz field-derivative-torque-induced nonlinear magnetization dynamics}

\author{Arpita Dutta}
\altaffiliation{arpita.dutta@niser.ac.in}
\affiliation{School of Physical Sciences, National Institute of Science Education and Research, An OCC of HBNI, Jatni, 752 050 Odisha, India}

\author{Pratyay Mukherjee}
\affiliation{Department of Physics, Indian Institute of Technology (ISM) Dhanbad, 826 004 Dhanbad, India}

\author{Swosti P. Sarangi}
\affiliation{School of Physical Sciences, National Institute of Science Education and Research, An OCC of HBNI, Jatni, 752 050 Odisha, India}

\author{Somasree Bhattacharjee}
\affiliation{Department of Physics, Indian Institute of Technology (ISM) Dhanbad, 826 004 Dhanbad, India}

\author{Shovon Pal}
\altaffiliation{shovon.pal@niser.ac.in}
\affiliation{School of Physical Sciences, National Institute of Science Education and Research, An OCC of HBNI, Jatni, 752 050 Odisha, India}

\author{Ritwik Mondal}
\altaffiliation{ritwik@iitism.ac.in}
\affiliation{Department of Physics, Indian Institute of Technology (ISM) Dhanbad, 826 004 Dhanbad, India}

\date{\today}

\begin{abstract}
The traditional Landau-Lifshitz-Gilbert (LLG) equation has often delineated the linear and nonlinear magnetization dynamics, even at ultrashort timescales e.g., femtoseconds. In contrast, several other non-relativistic and relativistic spin torques have been reported as an extension of the LLG spin dynamics. Here, we explore the contribution of the relativistic field-derivative torque (FDT) in the nonlinear THz magnetization dynamics response applied to ferrimagnets with high Gilbert damping and exchange magnon frequency. Our findings suggest that the FDT plays a significant role in magnetization dynamics in both linear and nonlinear regimes, bridging the gap between the traditional LLG spin dynamics and experimental observations. We find that the coherent THz magnon excitation amplitude is enhanced with the field-derivative torque. Furthermore, a phase shift in the magnon oscillation is induced by the FDT term. This phase shift is almost 90$^{\circ}$ for the antiferromagnet, while it is almost zero for the ferrimagnet under our investigation. Analyzing the dual THz excitation and their FDT, we find that the nonlinear signals can not be distinctly observed without the FDT terms. However, the inclusion of the FDT terms produces distinct nonlinear signals which matches extremely well with the previously reported experimental results.
\end{abstract}

\maketitle

\section{Introduction}
Femtosecond (fs) and picosecond (ps) magnetization dynamics offer promising potential for future generations of memory technology. Numerous experiments have demonstrated that magnetic spins can be manipulated and switched at ultrafast timescales through irradiation with femtosecond lasers or THz pulses~\cite{Bigot1996,Stanciu2007,Kimel2007-Review,Koopmans2010,Radu2011,Wienholdt2012PRL,Kampfrath2011}. In particular, applying two THz pulses
the collective spin modes can be driven to a nonlinear regime and the dynamics arising from the strong correlation of the modes can be exploited~\cite{lu2017coherent,pal2021origin,zhang2024terahertzupconversion,blank2023empowering,zhang2024terahertz,zhang2024terahertzupconversion}. These experiments were conducted using pump-probe techniques in either reflection or transmission geometry for ferromagnetic, antiferromagnetic, and ferrimagnetic materials. The corresponding spin dynamics were computed using the well-known Landau-Lifshitz-Gilbert (LLG) equation~\cite{landau35,Gilbert1955}. The LLG framework includes two different spin torques, namely: (a) the field torque that accounts for the precessional motion of magnetic spins around an effective magnetic field, and (b) the transverse damping torque that delineates the dissipation of energy from the magnetic spins such that they can align with the effective field at the equilibrium. The latter is characterized by a dimensionless Gilbert damping parameter that accounts for the rate of energy dissipation.

Despite the tremendous success of the LLG equation of motion in the realm of magnetization dynamics, there are scenarios which calls for the revision of the equation, for example, cases where the relativistic torques become prominent at the ultrashort timescales. To this end, the LLG equation has been extended by incorporating the non-relativistic and relativistic spin torques~\cite{Mondal2018PRB,Li2022APL}. It has been established that the damping torque in the LLG equation originates through the relativistic effects, namely the spin-orbit coupling~\cite{hickey09,Mondal2016,Thonig_2014,kambersky70,Nagyfalusi2024,barati14,kunes02,SakumaJAP2015}. In addition to the damping torque, the spin transfer torques~\cite{Li2003STT,slonczewski96,Berger1996,Ralph2008}, spin-orbit torques~\cite{Shao2021-ck,ManchonRMP,Gambardella2011,Choi2020,Jungwirth2016,mondal2021terahertz}, inertial spin torque~\cite{Ciornei2011,MONDAL_Review,neeraj2019experimental,unikandanunni2021inertial} and field-derivative torque (FDT)~\cite{Mondal2019PRB,Blank2021THz,dutta2024experimentalobservationrelativisticfieldderivative} have been included in the extended LLG equation of spin dynamics. While the microscopic origin of such torques has been presented, the impact of FDT has not been thoroughly examined in the literature. 

The FDT accounts for the spin torque exerted by the derivative of the applied field pulse. In such a case, if the applied field is constant with time, the FDT vanishes. However, the FDT turns out to be extremely important for a spin system triggered by a THz pulse. The origin of the FDT traces back to the relativistic spin-orbit coupling~\cite{Mondal2016}. The Hamiltonian expressing such spin-orbit coupling can be written as $\mathcal{H}_{\rm SOC} \propto {\bf S} \cdot \left[{\bf E}_{\rm ext} \times {\bf p}-{\bf p}\times {\bf E}_{\rm ext}\right]$, where ${\bf S}$ is the spin angular momentum, ${\bf E}_{\rm ext}$ is the external field from the THz pulse and $\bf p$ is momentum of the electron. Using Maxwell's electrodynamic equations, the Hamiltonian can be recast as $\mathcal{H}_{\rm SOC} \propto {\bf S} \cdot \left[2{\bf E}_{\rm ext} \times {\bf p}-i\hbar \frac{\partial {\bf B}}{\partial t}\right]$. While calculating the magnetization dynamics, it is the last term that gives rise to the FDT~\cite{Mondal2016}. Note that the THz field couples to the magnetic spins via the non-relativistic Zeeman Hamiltonian that exerts Zeeman torque (ZT) on the spins. 

The effect of FDT has been investigated within the atomistic spin dynamics simulations for antiferromagnetic NiO and CoO~\cite{Mondal2019PRB}. A comparison of the excitation of coherent magnon modes was made with ZT and ZT+FDT. While the magnitude of the magnon oscillations is enhanced with the FDT, the phase is also shifted by $\pi/2$ in antiferromagnets. However, it has been suggested that the effect of the FDT is stronger in materials having higher Gilbert damping and magnon resonance frequency~\cite{Mondal2019PRB,Blank2021THz}. To our knowledge, the first experimental realization of FDT has been comprehended in ferrimagnets, such as $\text{Gd}_{3/2}\text{Yb}_{1/2}\text{BiFe}_5\text{O}_{12}$, that has higher Gilbert damping values even at room temperature~\cite{dutta2024experimentalobservationrelativisticfieldderivative}. The nonlinear magnon spectra obtained in the experiment could only be explained by incorporating the FDT into the LLG equations~\cite{dutta2024experimentalobservationrelativisticfieldderivative}. In fact, without the FDT term, the simulations produced a fuzzy magnon spectrum where the nonlinear responses were not observed distinctly. 

In this work, we computationally explore the linear and nonlinear magnetization dynamics in the same ferrimagnetic Fe-garnet as in Ref.~\cite{dutta2024experimentalobservationrelativisticfieldderivative} and elucidate the role of FDT in the two-sublattice ferrimagnets. Our results clearly demonstrate that the THz magnon oscillation amplitude is enhanced with the FDT. We explore how the magnon oscillations depends on the ratio of unit cell volumes per spin, corresponding to the rare-earth and the Fe sublattices, involved in the FDT term. Whenever this volume ratio is exactly 1, the phase shift introduced by the FDT is approximately $\pi/2$, which agrees with the previously reported results~\cite{Mondal2019PRB}. A phase shift of $\pi$ can, however, be observed in the magnon oscillations (FDT incorporated) when the volume ratio is either greater than or less than 1. The phase shift in the magnon response induced by the FDT remains almost constant even if the ratio of volumes deviates further away from 1. We have also computed the nonlinear magnetization dynamics obtained via the dual THz excitation and the associated FDT terms. By analyzing the 2D spectral amplitudes of the nonlinear magnetization dynamics, we find that the FDT provides disentangled and distinct nonlinear signals that beautifully reproduces with the experimental observation~\cite{dutta2024experimentalobservationrelativisticfieldderivative}. Further, we investigated the effect of Gilbert damping where we found an upper limit of the Gilbert damping for obtaining distinct nonlinear signals, above which the system nonlinearities tend to become indiscernible.

The rest of the paper is organized as follows: Sec.~\ref{Sec2} outlines the spin model and our approach toward the simulation of the LLG equation incorporating the FDT term. The ferrimagnetic system that we study is discussed in Sec.~\ref{Sec3}. The results for the single THz pulse excitation are presented and discussed in Sec.~\ref{Sec4}A. The results for dual THz pulse excitation and the associated nonlinear magnetization dynamics are detailed in Sec.~\ref{Sec4}B. Finally, we conclude in Sec.~\ref{Sec5}.      

\section{Spin Model for Two-sublattice systems}
\label{Sec2}
The LLG spin dynamics can be represented by a combination of two terms: (a) the precession of magnetization vector ${\bf M}$ around an effective field ${\bf B}^{\rm eff}$, and (b) the transverse damping characterized by the dimensionless Gilbert damping parameter $\alpha$. These two terms put together comprise the following LLG spin dynamics 
\begin{align}
\label{Eq1}
\frac{{\rm d}{\bf M}_{i}}{{\rm d}t} = -\frac{\gamma_i}{1+\alpha_i^2} \,{\bf M}_{i} \times \left[{\bf B}_{i}^{\rm eff} + \frac{{\alpha_i}}{\vert {\bf M}_{i}\vert}\left({\bf M}_i\times
   {\bf B}_{i}^{\rm eff}\right)\right],
\end{align} 
where $i$ denotes the lattice indices and $\gamma = 28$ GHz/T is the gyromagnetic ratio. The effective magnetic field for each sublattice $\textbf{B}^{\rm eff}_i$ is calculated using the free-energy density of the system. The incorporation of the FDT term within the LLG spin dynamics changes the effective field to \cite{Mondal2016,Blank2021THz} 
\begin{equation}
   \textbf{B}_{i}^{\rm eff} \rightarrow\left(\textbf{B}_{i}^{\rm eff}-\frac{\alpha_i a^3_i}{\gamma_i\mu_{\rm B}}\frac{d\textbf{H}_{\rm THz}}{dt}\right),
    \label{Eq2}
\end{equation} 
where $a_{i}^{3}$ is the unit cell volume per spin for each sublattice and $\mu_{\rm B}$ is the Bohr magneton. Note that, in contrast to antiferromagnets, the contribution of the effective field in the presence of FDT is distinctly different for each sublattice in ferrimagnets. For most antiferromagnets like $\rm NiO$ or $\rm YFeO_{3}$, the magnetic moment originates from the same magnetic ion having equal sublattice spin volumes. Whereas, for ferrimagnets the contribution of the magnetic moment comes from disparate ions possessing non-identical sublattice spin volumes. Therefore, the effect of FDT is particularly interesting in ferrimagnets. This study is mostly focused on theoretical exploration, in continuation with our earlier experimental demonstration~\cite{dutta2024experimentalobservationrelativisticfieldderivative}, the linear and nonlinear THz magnetization dynamics in ferrimagnets under the influence of FDT. 

\begin{figure*}[t!]
    \centering
\includegraphics[width=0.95\linewidth]{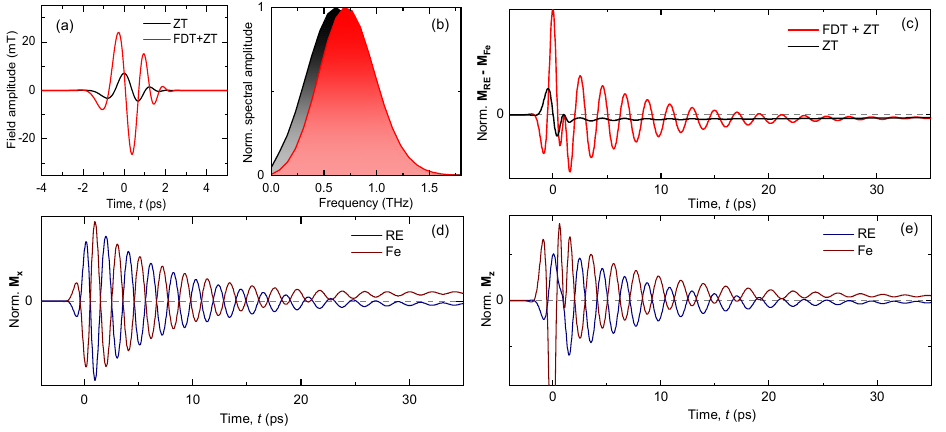}
    \caption{(a) Modelled effective field representing the THz-induced ZT and the total field incorporating FDT along with ZT. (b) The normalized spectra obtained by performing a fast Fourier transform of the fields in (a). The spectra show a clear shift in the frequency with the incorporation of FDT. (c) The exchange dynamics between the rare earth and the iron moments with and without the FDT. (d) The $x$- and (e) the $z$-components of the individual sublattice magnetization dynamics in the presence of FDT.}
    \label{fig1}
\end{figure*}

The free energy density $\Phi(\mathbf{M}_{\rm Fe} , \mathbf{M}_{\rm RE} )$ addressing a ferrimagnetic system can be represented as
\begin{align}
\label{Eq3}
   & \Phi(\mathbf{M}_{\rm Fe} , \mathbf{M}_{\rm RE}) = \nonumber\\
   & - \lambda \mathbf{M}_{\rm Fe} \cdot \mathbf{M}_{\rm RE} - \mathit K_{\rm Fe} \frac{\left(\mathbf{M}_{\rm Fe}\cdot \mathbf{n}\right)^{2}}{|\mathbf{M}_{\rm Fe}|^2} - \mathit K_{\rm RE} \frac{\left(\mathbf{M}_{\rm RE}\cdot \mathbf{n}\right)^{2}}{|\mathbf{M}_{\rm RE}|^2} \nonumber\\
    &- \mu_0 [\mathbf{H}_{\rm THz} (t, \tau) + \mathbf{H}_{\rm ext}] \cdot(\mathbf{M}_{\rm Fe}+\mathbf{M}_{\rm RE}) \nonumber\\
    &+ \frac{\mu_0}{2}\left(\mathbf{M}_{\rm Fe}\cdot \mathbf{n} + \mathbf{M}_{\rm RE}\cdot \mathbf{n}\right)^{2}, 
\end{align}
where the first term is the nearest neighbor magnetic exchange interaction between the rare-earth $(\textbf{M}_{\rm RE})$ and iron magnetizations $(\textbf{M}_{\rm Fe})$ with the Kaplan-Kittel exchange constant $\lambda$. The second and third terms represent the magnetic anisotropy energy for Fe and RE sublattices with uniaxial energy density $K_{\rm Fe}$ and $K_{\rm RE}$, respectively. The fourth term is the Zeeman interaction term with the THz magnetic field $(\textbf{H}_{\rm THz})$ and the external static magnetic field $(\textbf{H}_{\rm ext})$, and the last term is the demagnetization term with ${\bf n}$ representing the out-of-plane direction. In contrast to ferrimagnets like $\rm Tm_{3}Fe_{5}O_{12}$~\cite{Blank2021THz,blank2023effective}, where the demagnetization term was irrelevant, the demagnetization field in $\text{Gd}_{3/2}\text{Yb}_{1/2}\text{BiFe}_5\text{O}_{12}$ is comparable to the external magnetic field, which makes it crucial to incorporate in the system's free energy expression. Note that the intra-sublattice exchange interaction terms have not been included in the free energy density expression Eq.~(\ref{Eq3}). The reason is that our study is focused on the antiferromagnetic exchange mode that occurs at THz frequencies whereas the intra-sublattice exchange interactions are ferromagnetic-like. The effective magnetic field is then calculated by taking a derivative of the total energy with respect to the sublattice magnetization, i.e., $ \mathbf{{B}}_{i}^{\rm eff} = -\frac{\delta \Phi}{\delta \mathbf {M}_{i}}$. Using such an effective magnetic field, we solve the LLG equation, Eq.~(\ref{Eq1}), with and without incorporating FDT and thereafter compute the magnon response triggered by single and dual THz pulses.

\section{Application to rare-earth transition metal iron garnet}\label{Sec3}

To comprehend the implication of FDT in linear and nonlinear magnetization dynamics, we choose to work with a garnet having a formula unit $\rm R_{3}Fe_{5}O_{12}$ where $\rm R$ refers to a rare-earth element. In particular, we have opted for $\text{Gd}_{3/2}\text{Yb}_{1/2}\text{BiFe}_5\text{O}_{12}$ system because it possesses a high exchange resonance frequency and a substantial damping of 0.02 at room temperature~\cite{Satoh2012,Parchenko2013APL}. These properties of our chosen material make it perfect for inspecting the effects of FDT. Each unit cell of the garnet comprises three sublattices, one for the RE ion and the other two for the $\rm Fe$ ions. The net $\rm Fe$ magnetic moment assigned to the tetrahedral and octahedral sites shows antiferromagnetic coupling with the rare-earth moment in the dodecahedral site giving rise to a ferrimagnetic ordering at room temperature. The magnetic compensation temperature and the Curie temperature of the system are 96\,K and 573\,K respectively~\cite{Satoh2012, Parchenko2014}. In earlier studies~\cite{Kittel1960,Kaplan1953,dutta2024experimentalobservationrelativisticfieldderivative}, a high-frequency Kaplan-Kittel mode at around 0.4\,THz and a low-frequency backward volume magnetostatic wave (BVMW) mode at around 4.2\,GHz have been identified in the system by the non-thermal optical pump-probe technique via the inverse Faraday effect. The corresponding experiments were performed in the presence of an in-plane external magnetic field of 120\,mT. 

\begin{figure*}[t!]
    \centering
\includegraphics[width=0.75\linewidth]{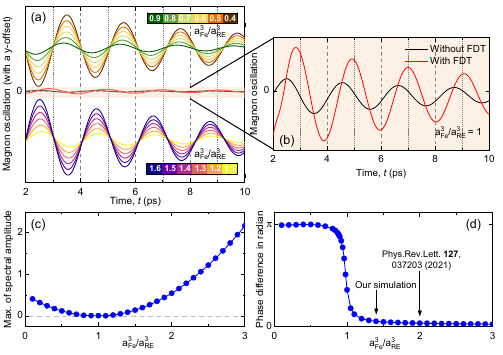}
    \caption{(a) THz excitation of the ferrimagnetic system in the absence and in the presence of field derivative torque (FDT) at different volume ratio of the iron and rare-earth sublattices. (b) The zoomed in part of (a) representing the difference in spin dynamics with and without FDT at $\rm a^{3}_{\rm Fe}=\rm a^{3}_{\rm RE}$. (c) The variation of the maxima of spectral amplitude obtained by the Fourier transforming the spectra in (a). (d) The difference in the phase of the magnon oscillations in (a) upon the incorporation of FDT when compared to the scenario without FDT.}
    \label{fig2}
\end{figure*}

To obtain these two co-existing magnon modes induced by impulsive excitation, we solve the two coupled LLG equations for the two sublattices (RE and Fe) and compute the magnetization dynamics for each sublattice. In addition we also compute the exchange dynamics via the N\'eel vector. These results are in excellent agreement with our experimental findings~\cite{dutta2024experimentalobservationrelativisticfieldderivative}. In our model, the free energy density has been evaluated using the parameters specified in Table~1 of Ref.~\cite{dutta2024experimentalobservationrelativisticfieldderivative}. Subsequently, we perform an in-depth analysis of the exchange dynamics induced by the resonant and coherent excitation by a single THz pulse and dual THz pulses, thereby, elucidating the significance of the FDT. 

To model the incident THz pulse, we use a single-cycle THz magnetic field, applied in the $x$-direction, obeying \cite{Mondal2019PRB} 
\begin{equation}
\label{Eq4}
    \mathbf {H}_{\rm THz}(t) = \rm H_{0} \, \cos(2\pi f_{0}\Gamma [\mathrm {e}^{t/\Gamma}-1]) \; \mathrm {e}^{-t^{2}/\sigma^{2}}\; \hat{\mathbf{x}}\,,
\end{equation}
where, $\rm f_{0}$ is the centre frequency, $\sigma$ is the pulse duration, $\rm H_{0}$ is the amplitude of the input THz magnetic field and $\Gamma$ is the chirp-time of the pulse. A typical THz field pulse and the field associated with the FDT term are shown in Fig.~\ref{fig1}(a) with $\sigma = 1$\,ps, ${\rm f_0} = 0.6$\,THz, $\Gamma = $ 2.6\,ps and ${\rm H}_0 = 7$\,mT. These parameters are chosen in such a way that the modelled pulses match with those used in our earlier experiments~ \cite{dutta2024experimentalobservationrelativisticfieldderivative}.  

\section{Results and discussions}
\label{Sec4}
\subsection{Effect of single THz pulse excitation}

We have studied the resonant and coherent THz pulse interactions in linear and nonlinear regimes in $\text{Gd}_{3/2}\text{Yb}_{1/2}\text{BiFe}_5\text{O}_{12}$. In Fig.~\ref{fig1}(a), we show the field amplitudes corresponding to the THz pulse that exerts ZT and the net field incorporating the field-derivative that exerts FDT, i.e., $\rm FDT+ZT$. While the amplitude of FDT is less than ZT, bringing in the FDT term with the ZT increases the overall amplitude of the induced field. In addition, and more importantly, the FDT term imparts a phase to the induced dynamics as shown in Fig.~\ref{fig1}(a). The normalized spectra obtained by performing the fast Fourier transform of the fields in Fig.~\ref{fig1}(a) are shown in Fig.~\ref{fig1}(b). Clearly, the spectra show a shift in the resonance frequency which can be attributed to the inclusion of the FDT. In the simulation, we also use a static magnetic field of 120\,mT along $y$-direction which saturates the magnetization and ensures that for each THz pulse excitation, the ground state remains the same. Fig.~\ref{fig1}(c) shows the THz-induced exchange (aka Kaplan-Kittel) dynamics of the ferrimagnetic system, denoted by the N\'eel vector $(\textbf{M}_{\rm RE}-\textbf{M}_{\rm Fe})$, both in presence and absence of FDT. Having introduced the FDT term, we notice that the system response, i.e., the amplitude of the exchange dynamics increases significantly while the phase remains the same in comparison to the scenario without FDT. The input fields, ZT and $\rm FDT+ZT$ in Fig.~\ref{fig1}(a), however, show an evident phase difference. The magnetization dynamics in the $x$ and $z$-component for the individual sublattices in the presence of FDT are shown in Figs.~\ref{fig1}(d) and~\ref{fig1}(e), respectively. We do not show the sublattice magnetization dynamics along the $y$-direction as that is the equilibrium direction. Along the $x$-direction, the magnon oscillations are equal in amplitude but opposite in phase in each sublattice. On the other hand, the oscillations in the $z$-component have different amplitudes indicating a different effective field along the $z$-axis for each sublattice. While the initial magnetization (before the THz interaction) along the $x$ and $z$ directions is zero, a finite magnetization is left after 30\,ps that results from the background BVMW mode at 4.2\,GHz. Note that in contrast to the experimental scenario, our simulations do not account for the absorption losses in our ferrimagnetic system.

Dealing with a ferrimagnetic system, it is important to note that the ratio of the unit cell volume per spin corresponding to the RE and Fe sublattices plays a crucial role in determining the significance of FDT underpinning the magnetization dynamics. The exchange dynamics driven magnon oscillation has been studied as a function of $\rm a^{3}_{\rm Fe}/\rm a^{3}_{\rm RE}$ in Fig.~\ref{fig2}(a). Here, $\rm a^{3}_{\rm Fe}$ and $\rm a^{3}_{\rm RE}$ are the unit cell volume per spin for Fe and RE sublattices, respectively. The FDT incorporated magnon oscillations show a notable increase in the amplitude except when the volume ratio is equal to 1. Fig.~\ref{fig2}(b) shows the N\'eel vector dynamics with and without FDT indicating almost a $\frac{\pi}{3}$ phase shift at $\rm a^{3}_{\rm Fe}/\rm a^{3}_{\rm RE} = 1$. This is consistent with the previously reported dynamics in antiferromagnetic NiO~\cite{Mondal2019PRB} except that the phase shift is not equal to $\frac{\pi}{2}$. This deviation stems from the unequal sublattice magnetizations in our model even when $\rm a^{3}_{\rm Fe} = \rm a^{3}_{\rm RE}$. In addition, the increase in magnon oscillation amplitude when $\rm a^{3}_{\rm Fe} = \rm a^{3}_{\rm RE}$ is much less as compared to other volume ratios [see Fig.~\ref{fig2}(a)].

\begin{figure}[t!]
    \centering
\includegraphics[width=\linewidth]{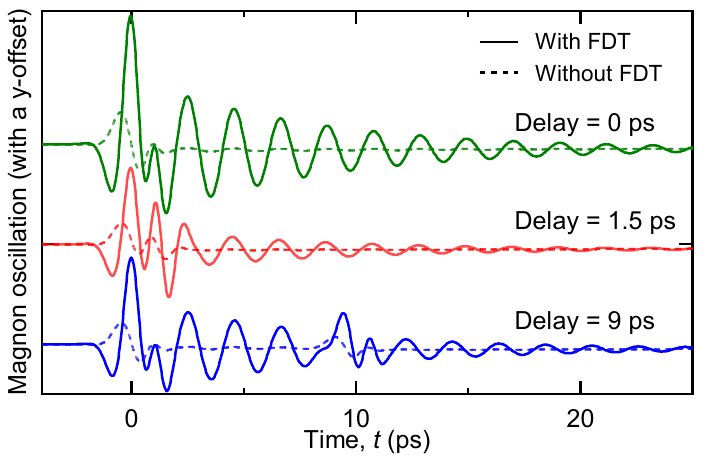}
    \caption{Exchange oscillations triggered by two THz pulses at three different delay times, 0\,ps, 1.5\,ps, and 9\,ps. The exchange dynamics in the presence and absence of FDT are shown by the solid and dashed lines, respectively.} 
    \label{fig3}
\end{figure}

Sweeping across different volume ratios, we observe that the amplitude of magnon oscillation and the phase depend on the volume ratio. In particular, as shown in Fig.~\ref{fig2}(a), when the ratio $\rm a^{3}_{\rm Fe}/\rm a^{3}_{\rm RE}<1$, a complete $\pi$ phase shift is obtained as compared to the dynamics without FDT. However, for $\rm a^{3}_{\rm Fe}/\rm a^{3}_{\rm RE}>1$, there is no phase shift when compared to the case without FDT. We have computed both the maximum spectral amplitude and the phase difference as a function of the ratio $\rm a^{3}_{\rm Fe}/\rm a^{3}_{\rm RE}$ in Figs.~\ref{fig2}(c) and ~\ref{fig2}(d), respectively~\cite{phase}. A noticeable dip in the maximum spectral amplitude near $\rm a^{3}_{\rm Fe}=\rm a^{3}_{\rm RE}$ hints to an underlying symmetry in the system corresponding to the nearly antiferromagnetic phase as described earlier. This also highlights that the effect of FDT is more pronounced in ferrimagnetic ordering, compared to the antiferromagnetic ones. The transition in the phase shift observed near the volume ratio $\rm a^{3}_{\rm Fe}/\rm a^{3}_{\rm RE} = 1$, along with the detailed volume analysis of the exchange dynamics, provide valuable insights in obtaining a defined phase in the THz magnetization dynamics, particularly when exploring new ferrimagnetic systems with different sublattices. 

In our investigations on the FDT-induced nonlinear dynamics (discussed in the next section) we considered the volume ratio to be 1.2. This is very close to a recent report~\cite{Blank2021THz} on the THz-induced exchange dynamics in ferrimagnetic $\rm Tm_{3}Fe_{5}O_{12}$, where the volume ratio $\rm a^{3}_{\rm Fe}/\rm a^{3}_{\rm Tm}$ was considered to be 2. From our systematic study, we also found that choosing a volume ratio greater than 1.2 does not change the obtained nonlinearities and no additional phase is introduced in the nonlinear magnon oscillations. 

\subsection{Effect of dual THz pulse excitation}
\begin{figure}[b!]
    \centering
\includegraphics[width=0.75\linewidth]{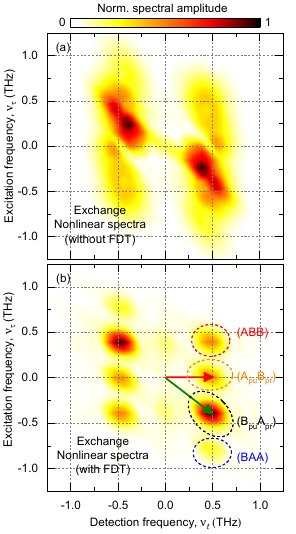}
    \caption{Normalized contour plot of 2D Fourier transformed spectra of the exchange nonlinear signal as a function of detection frequency and excitation frequency. (a) and (b) represent the spectra obtained by performing the numerical simulation without and with incorporating the FDT term, respectively. The spectra in (b) appear to have two intense pump-probe ($\rm A_{pu}-\rm B_{pr}$ and $\rm B_{pu}-\rm A_{pr}$) and two echo (ABB and BAA) signals. The green and red arrows indicate the frequency vectors corresponding to THz pulses A and B.}
    \label{fig4}
\end{figure}

\begin{figure*}[t!]
    \centering
\includegraphics[width=0.85\linewidth]{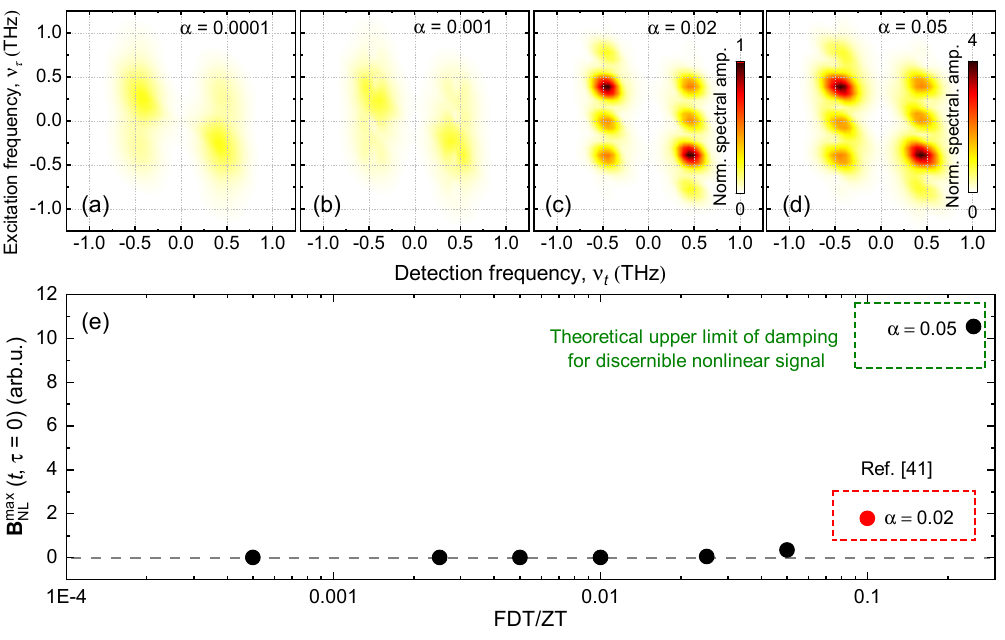}
    \caption{(a-d) The exchange nonlinear spectra of the Kaplan-Kittel mode for different orders of damping values in the ferrimagnetic system. (e) The maximum of the nonlinear signal, $\textbf{B}_{\rm NL}^{\rm max}$ showing a non-monotonic increase with the increase of the Gilbert damping parameter. The red point shows the result that corresponds to our earlier experimental condition~\cite{dutta2024experimentalobservationrelativisticfieldderivative}.}
    \label{fig5}
\end{figure*}

To analyze the effect of FDT in nonlinear THz magnetization dynamics, we consider the excitation of the ferrimagnet with two THz pulses, separated by a time delay $\tau$. Such dual THz excitation leads to the emission of several nonlinear signals. In our simulation, the effect of THz nonlinearity in spin dynamics has been accomplished by driving the exchange mode with two THz pulses, namely $\rm A$ and $\rm B$, where, the amplitude of the input magnetic field of THz pulse $\rm B$ is twice that of pulse $\rm A$. The shape of each THz pulse is governed by Eq.~(\ref{Eq4}). The interaction of each of these THz pulses with the sample is already discussed in the earlier section.

Figure~\ref{fig3} shows the time traces corresponding to the simultaneous interaction of both the THz pulses. Here only three delay points (i.e., for $\tau = 0, 1.5$ and $9$\,ps) are shown as exemplary cases. The figure also shows a dramatic difference in the time traces when FDT is incorporated (solid lines) as compared to the case when FDT is not included (dashed lines). Note that the THz pulse A is delayed by $\tau$, while THz pulse B excites the ferrimagnet without any delay. At $\tau = 0$\,ps, when the two THz pulses interact simultaneously with the ferrimagnetic system, we obtain the maximum enhancement of the exchange oscillations that persists for longer timescales up to 30\,ps. Evidently, the amplitude and phase of the exchange oscillations are modified according to the interference of the pulses governed by the delay between them. Once again, we observe that the FDT term is rather prominent in the dual THz excitation of the ferrimagnet. 

Next, the nonlinear exchange signal in the time domain $\textbf{B}_{\rm NL}(\textit{t},\tau)$ characterizes the nonlinear magnetization dynamics of the system. It is extracted by eliminating the single field responses, i.e., $\textbf{B}_{\rm A}(\textit{t},\tau)$ and $\textbf{B}_{\rm B}(\textit{t})$ from the two-field response $\textbf{B}_{\rm AB}(\textit{t}, \tau)$ according to the following equation
\begin{equation}
    \textbf{B}_{\rm NL}(\textit{t}, \tau) = \textbf{B}_{\rm AB}(\textit{t}, \tau) - \textbf{B}_{\rm A}(\textit{t}, \tau) - \textbf{B}_{\rm B}(\textit{t}).
\end{equation}
The 2D fast Fourier transform of the nonlinear signal in the time domain leads to the nonlinear spectra as a function of detection frequency ($\nu_{\textit{t}}$) and excitation frequency ($\nu_{\tau}$). The detection frequency refers to the specific frequency at which the exchange precessional motion occurs. In contrast, the excitation frequency provides information on the nonlinear frequency components as a result of the dual pulse THz excitation. While Fig.~\ref{fig4}(a) shows the nonlinear spectrum in the absence of FDT, Fig.~\ref{fig4}(b) shows the spectrum in the presence of FDT. Evidently, introducing the FDT term in our simulation leads to disentangled nonlinear signals whereas the absence of FDT results in a mere offset with no discernible nonlinear signal in the 2D spectrum. This establishes that not only the Zeeman field but also the derivative of the incident THz field has a significant effect in determining the nonlinear spin dynamics in a ferrimagnetic system with a high resonance frequency and simultaneously a high Gilbert damping parameter $\alpha$. Our simulations in Fig.~\ref{fig4}(b) are consistent with prior theoretical predictions~\cite{Mondal2019PRB} and experimental results~\cite{dutta2024experimentalobservationrelativisticfieldderivative}.

In the exchange nonlinear dynamics, we obtain two pump-probe signals namely, $\rm A_{pu}-\rm B_{pr}$ and $\rm B_{pu}-\rm A_{pr}$, and two echo signals, ABB and BAA. The pump-probe signals originate via two-field interaction from one of the THz pulse, followed by a single-field interaction from the other THz pulse. While the first THz pulse acts like a pump pulse to excite the exchange precession and the second THz pulse probes the ensuing dynamics. Here the interaction sequence of the THz fields is such that the information from the phase evolution of the probe fields survives, while the phase evolution of the pump fields vanish~\cite{pal2021origin}. The echo signals, in contrast, carry the phase information of both pulses. This is because before the complete dephasing of the exchange precession, excited by the first pulse, the second pulse interacts with the system and rephases the dynamics leading to a longer lifetime of the precessional motion. 

We now study the effect of Gilbert damping on the evolution of the nonlinear signal, as shown in Figs.~\ref{fig5}(a-d). Here, we change the damping, keeping the ${\rm f_0}$ fixed and obtain the nonlinear dynamics for a specified value of the unit cell volume ratio 1.41. The ratio of field-derivative torque to the Zeeman torque (FDT/ZT) is known to depend linearly on the Gilbert damping parameter $\alpha$ and the resonance frequency ${\rm f_0}$~\cite{Mondal2019PRB}. To scrutinize the effect of damping in our system, we plot the maximum of the nonlinear signal ($\textbf{B}_{\rm NL}^{\rm max}$) as a function of FDT/ZT in Fig.~\ref{fig5}(e). It is evident that above a particular threshold value of the damping, distinct nonlinear signals emerge. In the case of antiferromagnets like $\rm YFeO_{3}$ and NiO, $\alpha$ is very small, on the order of $10^{-3}$ to $10^{-4}$. In such a scenario, the effect of FDT for such antiferromagnets is negligibly small and the incorporation of FDT in the nonlinear magnetization dynamics does not introduce any distinct difference in the nonlinear magnon spectra. In contrast, as we increase the damping hypothetically in our system, the strength of the nonlinear signal increases, showing a non-monotonic behavior, see Fig.~\ref{fig5}(e). Note that above a very high damping parameter $\alpha\ge 0.05$, we enter into a regime of light-spin interaction where the magnon mode bifurcates (see Appendix \ref{AppendixA}) and the signals become indiscernible -- the microscopic picture of which is still under investigation and is not well understood.

\section{Conclusion}
\label{Sec5}
Using computational approach, we have investigated the consequences of the modified LLG equation that incorporates FDT in the linear and nonlinear magnetization dynamics of a ferrimagnetic system. From prior theoretical results, it is known that the effect of FDT is pronounced for high-frequency excitations and scales linearly with Gilbert damping of the system. Here we found that the high-frequency exchange magnon oscillation of the ferrimagnetic system exhibits an intriguing correlation between its amplitude and phase with the unit cell volumes per spin. Notably, this behavior showcases distinct dynamics that are in contrast to the dynamics observed in the antiferromagnetic NiO or ferrimagnetic $\rm Tm_{3}Fe_{5}O_{12}$. Incorporation of the FDT in the magnetization dynamics of our ferrimagnetic system promotes a large amplitude magnon oscillation with an additional phase compared to the dynamics with Zeeman torque only. Additionally, the nonlinear response of the exchange dynamics in the presence of FDT unveils distinct dependence on system parameters and at the same time corroborates with the previously reported experimental findings.\\

\begin{figure}[b!]
    \centering
\includegraphics[width=\linewidth]{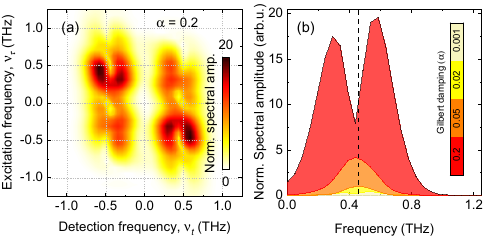}
    \caption{(a) Contour plot of the exchange nonlinear signal of the exchange mode at a hypothetical high damping of 0.2. The spectra appear deformed, where the nonlinear signals are indistinguishable. (b) The normalized spectral amplitudes of the nonlinear exchange mode at different Gilbert damping. The line scans are taken by a horizontal cut at $\nu_{\tau}=-0.4$\,THz from the nonlinear spectra, shown in (a) and Figs.~\ref{fig5}(b-d).}
    \label{fig6}
\end{figure}

\section*{Acknowledgments}
A.D. and S.P. acknowledge the support from DAE through the project Basic Research in Physical and Multidisciplinary Sciences via RIN4001. S.P. also acknowledges the start-up support from DAE through NISER and SERB through SERB-SRG via Project No. SRG/2022/000290. R.M. acknowledges SERB-SRG via Project No. SRG/2023/000612 and the faculty research scheme at IIT (ISM) Dhanbad, India under Project No. FRS(196)/2023-2024/PHYSICS. The authors acknowledge C. Tzschaschel for the fruitful discussions. 
\appendix
\section{Effect of large Gilbert damping on the nonlinear magnetization dynamics}
\label{AppendixA}
To exploit the nonlinear magnetization dynamics of the exchange mode at a very high damping value, we hypothetically increase the damping parameter to 0.2, keeping the magnon resonance frequency constant and obtain the nonlinear spectra shown in Fig.~\ref{fig6}(a). At such a high damping, the nonlinear magnon oscillations due to the exchange mode has diminished completely and the nonlinear spectra becomes deformed with indistinguishable nonlinear signals. The individual line scans at a particular excitation frequency of $\nu_{\tau}=-0.4$\,THz of the nonlinear spectra corresponding to different damping parameters, shown in Fig.~\ref{fig6}(b), highlights the emergence and enhancement of the nonlinear signals till a certain value of $\alpha$, above which magnetization dynamics deform, a regime which is still under investigation.  

\bibliography{Ref}

\begin{thebibliography}{48}
\bibitem [{pha()}]{phase}%
  \BibitemOpen
  \href@noop {} {}\bibinfo {note} {The phase shift is calculated as the difference in the phase with FDT and that without FDT. All the phase shift values are mapped to $[-\pi, \pi]$ and thereafter, we take the absolute of these values to get the phase shift in the interval $[0,\pi]$}\BibitemShut {NoStop}%
\end{thebibliography}

\end{document}